\documentclass[%
 reprint,
superscriptaddress,
 amsmath,amssymb,
 aps,
prb,
]{revtex4-2}

\usepackage{graphicx}
\usepackage{braket}
\usepackage{comment}

\begin{document}

\title{Quantitative Photoemission Predictions of Semiconducting Photocathodes from Many-Body \textit{Ab Initio} Theory}

\author{Richard Schier}
\email{richard.schier@uni-jena.de}
\affiliation{%
 Carl von Ossietzky Universit\"at Oldenburg, Institute of Physics, 26129 Oldenburg,
Germany
}%
\affiliation{Friedrich-Schiller Universit\"at Jena, Institute of Condensed Matter Theory and Optics, 07743 Jena, Germany}

\author{Chen Wang}
\affiliation{Helmholtz-Zentrum Berlin, 12489 Berlin, Germany}
\affiliation{Universit\"at Siegen, Institut f\"ur Werkstofftechnik, 57076 Siegen, Germany}

\author{Jonas Dube}
\affiliation{Helmholtz-Zentrum Berlin, 12489 Berlin, Germany}
\affiliation{Humboldt-Universit\"at zu Berlin, Institut f\"ur Physik, 12489 Berlin, Germany}

\author{Julius K\"uhn}
\affiliation{Helmholtz-Zentrum Berlin, 12489 Berlin, Germany}

\author{Alice Galdi}
\affiliation{Helmholtz-Zentrum Berlin, 12489 Berlin, Germany}
\affiliation{Universit\'a degli Studi di Salerno, Dipartimento di Ingegneria Industriale, 84084 Fisciano (SA), Italy}

\author{Thorsten Kamps}
\affiliation{Helmholtz-Zentrum Berlin, 12489 Berlin, Germany}
\affiliation{Humboldt-Universit\"at zu Berlin, Institut f\"ur Physik, 12489 Berlin, Germany}

\author{Caterina Cocchi}
\email{caterina.cocchi@uni-jena.de}
\affiliation{%
 Carl von Ossietzky Universit\"at Oldenburg, Institute of Physics, 26129 Oldenburg,
Germany
}%
\affiliation{Friedrich-Schiller Universit\"at Jena, Institute for Condensed Matter Theory and Optics, 07743 Jena, Germany}

\date{\today}

\begin{abstract}
The development of high-performance electron sources requires theoretical frameworks that accurately link the microscopic electronic properties of cathode materials to their macroscopic photoemission observables. Here, we present a many-body extension of the three-step photoemission model for semiconducting photocathodes, directly integrating the $GW$ approximation and the solution of the Bethe-Salpeter equation on top of density functional theory (DFT). This approach overcomes the intrinsic limitations of standard DFT by explicitly accounting for quasiparticle and excitonic effects in the photoexcitation process. The quantum efficiency (QE) is evaluated by combining the \textit{ab initio} absorption with an emission probability derived as an exciton-weighted average. We validate this model on representative alkali antimonides and demonstrate that a qualitative many-body description successfully captures complex spectral features that empirical models fail to reproduce. Furthermore, by incorporating macroscopic optical effects such as thin-film interference and polarization via Fresnel post-processing, we achieve quantitative agreement with experimental QE values without any adjustment. Minor discrepancies near the photoemission threshold are attributed to the idealized surface barrier adopted in the model and impurity effects in the samples, highlighting specific directions for future refinements. This work establishes a robust, parameter-free \textit{ab initio} tool that bridges microscopic electronic correlation with macroscopic observables, providing a critical pathway for the rational design of next-generation electron sources.
\end{abstract}

\maketitle


\section{Introduction}
The simulation of photoemission from solids has become an active research field in the last few years, driven by the need for high-quality electron sources in advanced applications such as free-electron lasers and particle accelerators~\cite{fili+22rmp,scha+23jmcc}. Over the last several decades, photoemission models have evolved significantly, moving from foundational phenomenological frameworks~\cite{spic58pr,kane62pr,berg-spic64pr} to sophisticated first-principles schemes~\cite{cami+16cms,anto+20prb,saha+22apl}. The pioneering work by Berglund and Spicer in 1964~\cite{berg-spic64pr} established the so-called three-step model, which conceptualizes photoemission in three sequential processes: (i) optical excitation generating photoelectrons, (ii) their scattering and transport to the surface, and (iii) their emission into the vacuum. Despite the success of the three-step model and its variants, particularly for polycrystalline or disordered cathodes~\cite{dimi+17jap,saha+23jap}, modern applications require more sophisticated methods accounting for the electronic structure of the materials and the complex physics of thin-film heterostructures.

The increasing demand for high-quality photoemission predictions has boosted the development of parameter-free approaches that can be applied to any material, independently of their synthesis and characterization. The one-step model specifically developed for metals~\cite{kark+17prb} treats photoemission as a single quantum mechanical process. The total transition rate is calculated using Fermi's golden rule, offering a quantitative explanation for the dependence of quantum efficiency (QE) on the polarization and angle of incidence of the incoming radiation. Inelastic collisions undergone by energetic electrons are explicitly incorporated via an effective scattering length. This model has been successfully applied to evaluate the emission from various metallic surfaces~\cite{kark+17prb,cami+16cms,adhi+19aipa} even under \textit{operando} conditions~\cite{schr-gowr19njp}.

The model proposed by Jensen and coworkers in the early 2000s~\cite{jens+06apl} aimed at improving the description of the intrinsic emittance of metal photocathodes beyond simple analytical methods. This framework has been subsequently extended to treat semiconductors, explicitly including band bending, phonon scattering mechanisms, effective mass differences, and the characteristic absence of electrons in the conduction band at rest conditions~\cite{jens+08jap}. This model was successfully applied to Cs$_3$Sb~\cite{jens+08jap,jens+10prab}, an established semiconducting photocathode for particle accelerators~\cite{cult+11apl,parz+23aplm,owus+25apl}, and to its ternary sibling CsK$_2$Sb~\cite{jens+08prab}.

The integration of photoemission models into \textit{ab initio} workflows has substantially enhanced their predictive power. In the scheme proposed by Antoniuk et al.~\cite{anto+20prb}, the electronic structure is computed from density-functional theory (DFT), and the emission probability is determined using Fermi's golden rule, neglecting scattering effects. A subsequent recipe proposed by Nangoi et al.~\cite{nang+21prb} enhances the description of the second step by incorporating coherent electron-phonon scattering. These DFT-based photoemission models enable efficient material screening~\cite{anto+21am,schi+24afm} and a systematic assessment of structural or chemical modifications even in complex geometries such as heterostructures and alloys~\cite{chan+18prb,napi+19prap}. 

On the downside, these pure DFT-based approaches suffer from crucial limitations in reproducing the electronic structure and the optical properties of semiconducting photocathodes. Relying on semi-local exchange-correlation functionals and on the independent-particle approximation for evaluating optical transitions, they treat particle interactions in a mean-field fashion, neglecting electron-electron and electron-hole correlations that are usually vital for a reliable description of photoemission. The adopted simplifications not only require \textit{ad hoc} shifts of the electronic bands to match experiments~\cite{anto+20prb} but also overshadow the fundamental many-body nature of the photoexcitation process in semiconductors~\cite{cocc+19jpcm,amad+21jpcm,cocc-sass21micromachines,schi-cocc25prm,xu+25es}.

In this work, we provide a many-body extension of the DFT-based three-step model for photoemission, integrating perturbation theory techniques such as the $GW$ approximation for the electronic self-energy and the Bethe-Salpeter equation to describe (bound) excitons. Adopting an energy-dependent transmission probability, we derive the spectral response for a series of (multi-)alkali antimonide semiconductors and benchmark our predictions against experimental quantum efficiency measurements. While the output of this workflow already contains all qualitative features probed in experiments, quantitative predictions can be achieved from a Frenel-based supervised post-processing of the emission function, including macroscopic parameters such as sample thickness and incidence angle. Our findings bridge the gap between microscopic electronic structure calculations and macroscopic photoemission observables, providing an invaluable tool for the rational optimization of photocathode materials for particle accelerators.

\section{Methods}
Before introducing the proposed many-body extension of the \textit{ab initio} three-step photoemission model, we review the formalism of DFT~\cite{hohe-kohn64pr} and many-body perturbation theory (MBPT)~\cite{onid+02rmp} as the fundamental basis of our development (Sec.~\ref{ssec:background}). This section is mainly addressed to an audience that is not familiar with these approaches and can be skipped by experienced readers. In Sec.~\ref{ssec:exp-details}, we provide the experimental settings employed to prepare the photocathode samples and to measure their photoemission yield.

\subsection{Theoretical Background}
\label{ssec:background}
In the Kohn-Sham (KS) implementation of DFT~\cite{kohn-sham65physrev}, the fully interacting many-body problem is mapped by a fictitious system of independent particles, ruled by the Schr\"odinger-like equation:
\begin{equation}
   \hat{h}^{KS} \ket{\phi_{n\text{k}}} = \epsilon_{n\text{k}}^{KS} \ket{\phi_{n\text{k}}}.
\end{equation}
The KS Hamiltonian
\begin{equation}
    \hat{h}^{KS} = \hat{t} + \hat{v}^{\text{eff}} = \hat{t} + \hat{v}^{\text{ext}} + \hat{v}^{\text{H}} + \hat{v}^{\text{xc}}
    \label{eq:KS-Ham}
\end{equation}
includes the single-particle kinetic energy operator $\hat{t}$ and an effective potential $\hat{v}^{\text{eff}}$, which consists of the external potential ($\hat{v}^{\text{ext}}$), accounting for the electron-nuclear attraction, the Hartree potential ($\hat{v}^{\text{H}}$), capturing the classical repulsion experienced by an electron in a negative charge distribution, and the exchange-correlation (xc) potential ($\hat{v}^{\text{xc}}$), embedding electronic interactions beyond the classical picture. Since the exact form of $\hat{v}^{\text{xc}}$ is unknown, it must be approximated. The quality of this approximation ultimately determines the accuracy of the resulting electronic structure. While semi-local functionals are known to dramatically underestimate the fundamental gap of solids even by a factor of 2, hybrid functionals mitigate this issue but at substantially higher computational costs~\cite{sass-cocc21es}, which become unaffordable for large and complex systems or for screening an extended configurational space. The meta-generalized gradient approximation (meta-GGA), as implemented in the r$^2$SCAN functional~\cite{sun+15prl}, represents a reasonable compromise between accuracy and computational costs, especially for photocathode materials~\cite{sass-cocc21es,schi+22prm}. However, the numerical instabilities that are still present in many implementations require special care in high-throughput applications~\cite{sass-cocc24npjcm,schi+24ats}. 

The $GW$ approximation~\cite{hedi65:physrev} is the state-of-the-art method to calculate the electronic structure of solids, overcoming the infamous ``band-gap problem'' of DFT~\cite{rein18wircms}. In the perturbative $G_0W_0$ approach, the electronic self-energy is calculated as
\begin{equation}
\Sigma (r, r', \omega) = \frac{i}{2\pi} \int G_0 (r,r',\omega + \omega') W_0(r, r', \omega') e^{i \omega' \eta} d\omega',
\label{eq:Sigma}
\end{equation}
where $G_0$ is the single-particle Green's function computed on top of DFT and $W_0$ is the Coulomb potential screened by the frequency-dependent dielectric function of the material, $\epsilon (\omega)$. The self-energy computed from Eq.~\eqref{eq:Sigma} enters the quasi-particle (QP) equation for the electron energies,
\begin{equation}
\epsilon_{n\text{k}}^{\text{QP}} = \epsilon_{n\text{k}} + Z_{n\text{k}} [\Re\Sigma_{n\text{k}} (\epsilon_{n\text{k}}) - V_{n\text{k}}^{\text{xc}}],
\label{eq:QP}
\end{equation}
where the renormalization factor $Z_{n\text{k}}$ accounts for the energy-dependence of the self-energy from which the contribution from the xc potential, $V_{n\text{k}}^{\text{xc}} = \langle \phi_{n\text{k}}|\hat{v}^{\text{xc}}|\phi_{n\text{k}} \rangle$, must be subtracted.

To calculate optical excitations including excitonic effects, the equation of motion of the electron-hole correlation function, known as the Bethe-Salpeter equation (BSE)~\cite{salp-beth51physrev}, must be solved. In the context of electronic-structure simulations, the BSE is mapped into the eigenvalue problem
\begin{equation}\label{eq:BSE}
\sum_{v^{\prime} c^{\prime} \mathbf{k}^{\prime}} H_{v c \mathbf{k}, v^{\prime} c^{\prime} \mathbf{k}^{\prime}}^{\mathrm{BSE}} A_{v^{\prime} c^{\prime} \mathbf{k}^{\prime}}^\lambda=E^\lambda A_{v c \mathbf{k}}^\lambda,
\end{equation}
where the indices $v$ and $c$ label valence and conduction states, respectively. The two-particle BSE Hamiltonian 
\begin{equation}
\hat{H}^{\text{BSE}} = \hat{H}^{\text{diag}} + 2 \hat{H}^{\text{x}} + \hat{H}^{\text{c}}
\label{eq:H_BSE}
\end{equation}
consists of three terms. The diagonal term ($\hat{H}^{\text{diag}}$) describes vertical electronic transitions, the exchange term ($\hat{H}^{\text{x}}$), which is multiplied by 2 assuming spin-degenerate systems, accounts for the repulsive exchange interaction within the fermionic electron-hole pairs, while the direct term ($\hat{H}^{\text{c}}$) includes the attractive electron-hole screened Coulomb interaction. 
The diagonalization of Eq.~\eqref{eq:BSE} delivers excitation energy eigenvalues $E^{\lambda}$ and eigenvectors $A^{\lambda}$, which contain information about the oscillator strength and composition of the $\lambda$-th excited state. $A^{\lambda}$ enter the expression of the imaginary part of the macroscopic dielectric tensor, 
\begin{align}\label{eqn:macrdielectric}
\Im \epsilon_M=\frac{8 \pi^2}{\Omega} \sum_\lambda\left|\mathbf{t}^\lambda\right|^2 \delta\left(\omega-E^\lambda\right),
\end{align}
through the transition coefficients 
\begin{equation}\label{eqn:transitioncoeffs}
\mathbf{t}^\lambda=\sum_{v c \mathbf{k}} A_{v c \mathbf{k}}^\lambda \frac{\langle \phi_{v\text{k}}|\hat{\mathbf{p}}| \phi_{c\text{k}}\rangle}{\epsilon_{c \mathbf{k}}^{\rm QP}-\epsilon_{v \mathbf{k}}^{\rm QP}}.
\end{equation}
The momentum matrix elements in the numerator of Eq.~\eqref{eqn:transitioncoeffs} couple transitions between occupied ($v$) and unoccupied ($c$) KS states. The ideally infinite number of conduction states assumed in the sum is numerically converged to a finite value. Finally, the contribution of specific single-particle transitions to the electron-hole state is quantified by the so-called \textit{exciton weights} for holes 
\begin{equation}
w^{\lambda}_{v\mathbf{k}} = \sum_c |A^{\lambda}_{vc\mathbf{k}}|^2
\label{eq:w_h}
\end{equation}
and electrons
\begin{equation}
 w^{\lambda}_{c\mathbf{k}} = \sum_v |A^{\lambda}_{vc\mathbf{k}}|^2.
\label{eq:w_e}
\end{equation}

\subsection{Experimental Setup}
\label{ssec:exp-details}
The experimental data for the ternary alkali antimonides were collected in the photocathode laboratory at Helmholtz-Zentrum Berlin für Materialien und Energie GmbH (HZB)~\cite{dube+25jap}. The photocathodes are prepared on a Mo substrate to meet the high thermal and electrical requirements necessary for operation in a superconducting radio-frequency accelerator. The Na-K-Sb sample TWH25 (Fig.~\ref{fig:results_ternary}a) was prepared by simultaneous deposition of all three materials (triple evaporation) from thermal effusion cells~\cite{wang+26jpd}, while the Cs-K-Sb photocathode G002 (Fig.~\ref{fig:results_ternary}b) was grown from deposition of a pure Sb layer and subsequent co-deposition of the alkali metals from dispensers~\cite{schm+18prab, cocc+19sr}. These samples were prepared, stored, and characterized under ultra-high vacuum conditions to avoid any degradation of the photocathodes due to residual gases. During preparation, the QE was used as feedback to optimize the growth parameters~\cite{dube+25jap,wang+26jpd}.

To perform spectrally resolved QE measurements, the photocathodes were illuminated by a monochromatic light source, consisting of a Xenon lamp and a monochromator. The photocurrent was measured by a picoammeter attached to a copper wire approximately 10~mm away from the photocathode surface. The wire was positively biased and acted as a pick-up anode~\cite{kirs+17ipac}. Under these conditions, the QE was determined by:
\begin{equation}
    \text{QE} = \frac{I_e hc}{P_\gamma\lambda e},
\end{equation}
where $I_e$ is the measured current, $P_\gamma$ the optical power, and $\lambda$ the wavelength of the incoming radiation.

\section{Photoemission Model from Ab initio Many-Body Theory}
\label{sec:model}
The photoemission scheme from \textit{ab initio} many-body theory proposed in this work is based on the three-step model originally formulated by Berglund and Spicer~\cite{spic58pr,berg-spic64pr}. In this framework, the photoemission process is decomposed into three sequential steps: First, an incident photon excites an electron from an initially occupied valence state to the conduction band. In the second step, the excited electron travels through the crystal toward the surface, potentially undergoing scattering events with other electrons and/or with phonons. In the third and final stage, the electron overcomes the surface potential barrier and is emitted.

\begin{figure}
	\centering
	\includegraphics[width=0.48\textwidth]{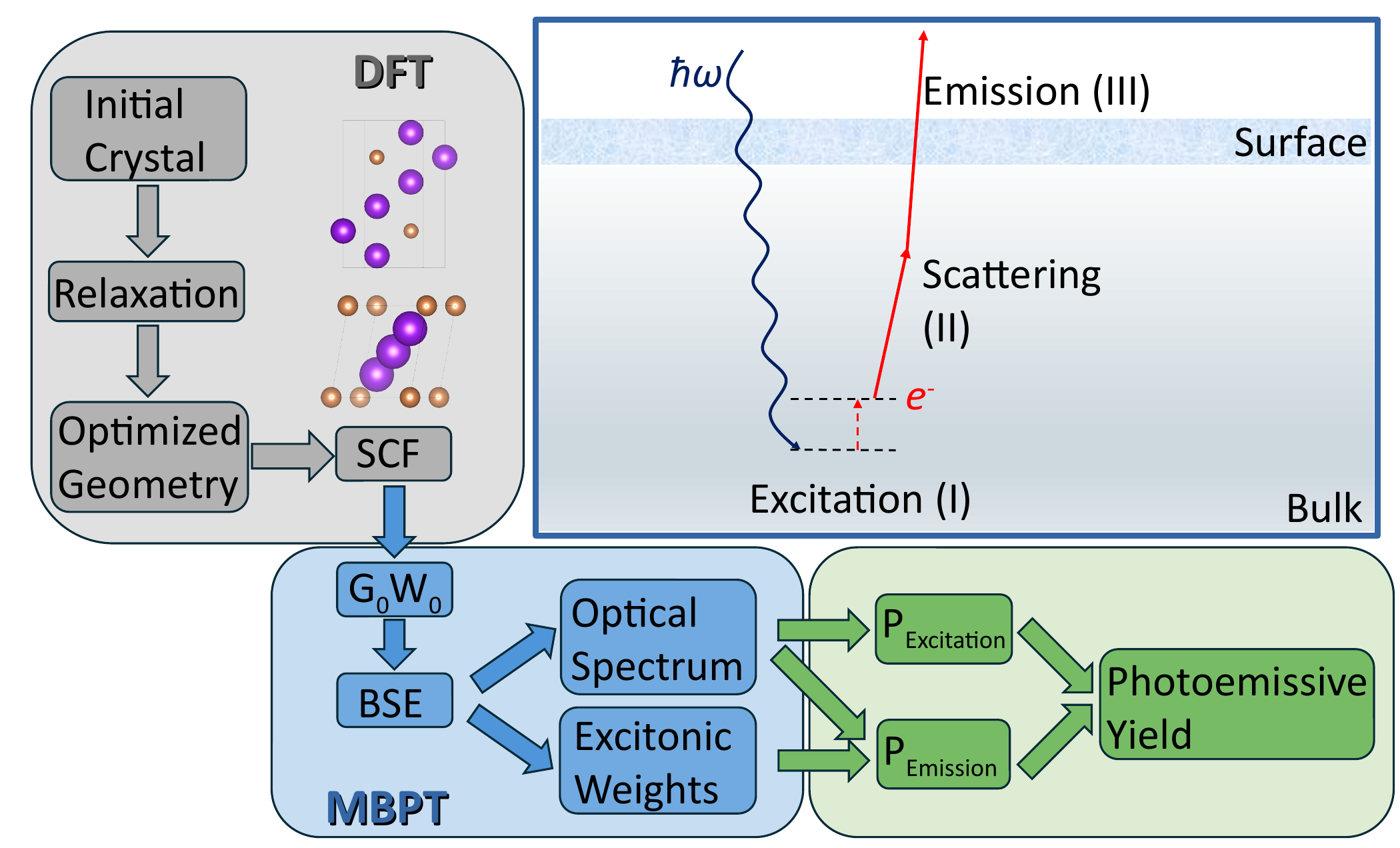}
\caption{Schematic overview of the three-step model for photoemission implemented from \textit{ab initio} many-body theory. The colored blocks illustrate each computational step, including DFT calculations of the ground-state properties (grey), MBPT runs ($G_0W_0$+BSE) to access excited state properties (blue), and final post-processing to compute the photoemission yield (green). 
}
	\label{fig:workflow}
\end{figure}

The proposed model implements the above-mentioned steps in the framework of DFT and MBPT, see Fig.~\ref{fig:workflow}. The starting point is the definition of the crystal structure, either tabulated or retrieved from computational or experimental databases. In the latter scenario, \textit{ad hoc} routines embedded in the in-house developed \texttt{Python} library \texttt{aim$^2$dat}~\cite{aim2dat} offer an effective and well-tested tool for data mining~\cite{schi+24ats}. Following the input setup, the structures are relaxed, including both lattice optimization and interatomic force minimization, as discussed in previous work~\cite{schi-cocc25prm,xu+25es}. The final DFT step includes a self-consistent field calculation as a starting point for MBPT, including both the QP correction to the electronic structure from $G_0W_0$ (Eq.~\ref{eq:QP}) and the solution of the BSE to compute the optical spectrum (Eq.~\ref{eqn:macrdielectric}) and excitonic weights (Eqs.~\ref{eq:w_h} and \ref{eq:w_e}). The $G_0W_0$+BSE results are post-processed to evaluate the excitation and emission probability, finally delivering the QE as discussed in detail below.
To illustrate the physical significance and the main ingredient entering each step, we take as an example hexagonal K$_3$Sb, an alkali antimonide crystal thoroughly characterized from DFT and MBPT in previous work~\cite{schi-cocc25prm}.

\subsection{Step 1: Photoexcitation}
The excitation probability quantifies the first step of photoemission. The standard expression of the imaginary part of the macroscopic dielectric function obtained by solving the BSE (Eq.~\ref{eqn:macrdielectric}) is conveniently reformulated as follows~\cite{vorw+19es}:
\begin{equation}\label{eq:epsilon_Lorentz}
\Im \epsilon_M = \frac{4 \pi^2}{\pi \Gamma \Omega} \sum_{\lambda} |\mathbf{t}_{\lambda}|^2 \left(\frac{\Gamma^2}{(\omega - E_{\lambda})^2 + \Gamma^2} - \frac{\Gamma^2}{(\omega + E_{\lambda})^2 + \Gamma^2}\right).
\end{equation}
In contrast to the general form in Eq.~\eqref{eqn:macrdielectric}, where the electron excitation is modeled by a $\delta$-function, Eq.~\eqref{eq:epsilon_Lorentz} includes a finite Lorentzian broadening $\Gamma$ and explicitly decouples excitation and de-excitation processes. $E_{\lambda}$ and $\mathbf{t}_{\lambda}$ are the excitation energies and transition coefficients defined in Eq.~\eqref{eq:BSE} and Eq.~\eqref{eqn:transitioncoeffs}, respectively.

\begin{figure}
	\centering
	\includegraphics[width=0.5\textwidth]{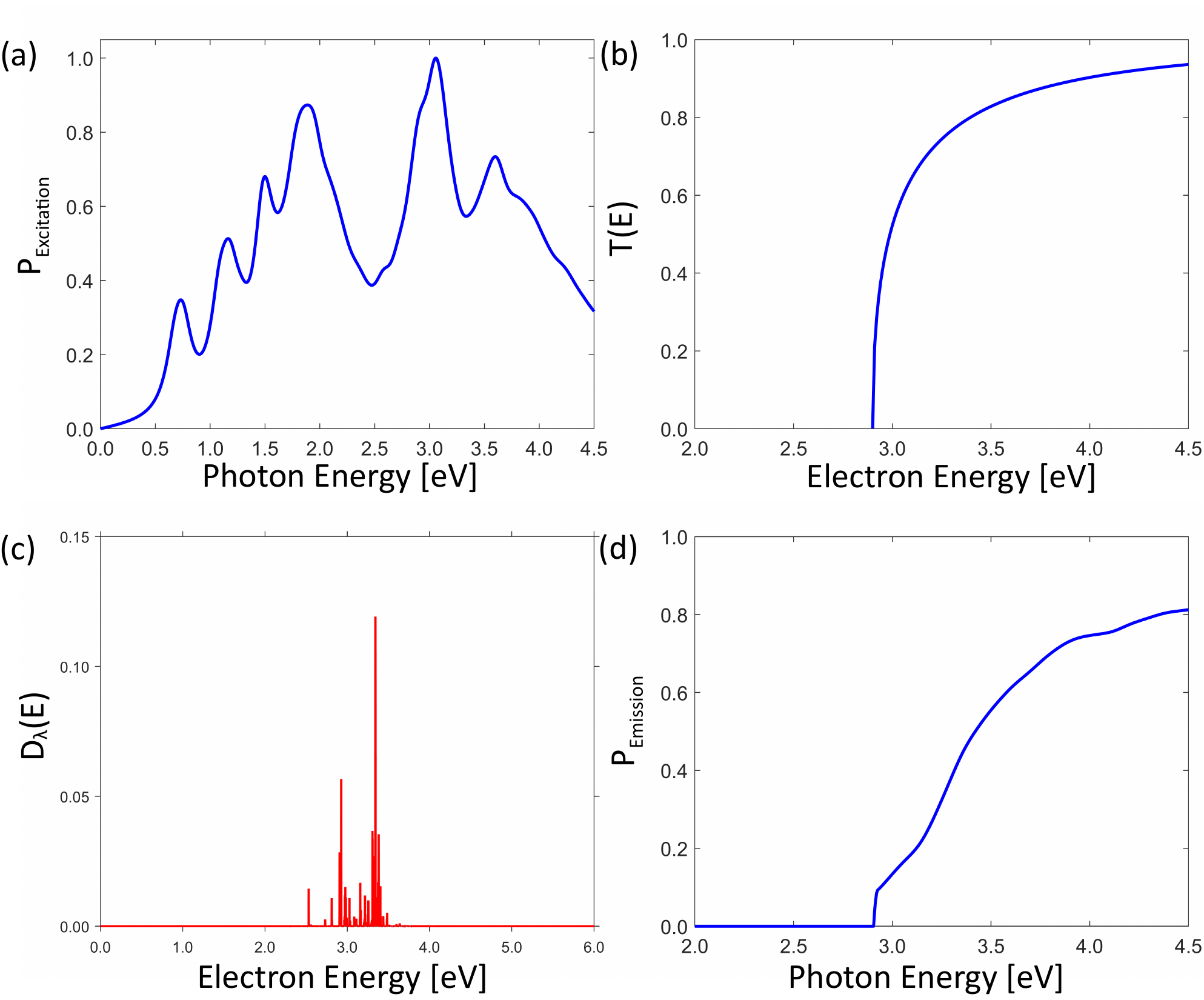}
\caption{Results for K$_3$Sb: (a) Optical absorption averaged over components~\cite{schi-cocc25prm}, (b) probability of emission depending on the energy of the excited electron, with onset from experiment~\cite{spic58pr}, (c) exemplary energy distribution of an excited electron at 3.6~eV, and (d) total calculated emission probability by photon energy.
}
	\label{fig:plots_K3Sb}
\end{figure}

The macroscopic dielectric function defined in Eq.~\eqref{eq:epsilon_Lorentz} is a tensor, with the number of inequivalent components and their magnitude dictated by crystal symmetry. While all the elements of $\Im \epsilon_M (\omega)$ are obtained by diagonalizing Eq.~\eqref{eq:BSE}, experiments probing photocathode emission are typically insensitive to the polarization direction. To mimic this scenario and assuming that off-diagonal contributions are absent or negligible, we average over all diagonal components and define the photoexcitation probability as the trace of the imaginary part of the dielectric function normalized to its absolute maximum in the considered spectral region:
\begin{equation}
    P_{\text{Excitation}} (\omega) = \dfrac{\ \Im \epsilon_M^{\text{avg}} (\omega)}{\text{max}_{\omega} [\Im \epsilon_M^{\text{avg}} (\omega)]} .
    \label{eq:p_exc}
\end{equation}
The excitation probability of hexagonal K$_3$Sb, calculated from Eq.~\eqref{eq:p_exc} including both in-plane and out-of-plane contributions from the dielectric tensor~\cite{schi-cocc25prm}, is shown in Fig.~\ref{fig:plots_K3Sb}(a). 

\subsection{Step 2: Electron Transport and Scattering}
Electron transport and scattering events included in the second step of Spicer's model are treated here as elastic processes, neglecting any dissipation effects. This assumption, already adopted by Antoniuk et al.~\cite{anto+20prb}, is justified both by material properties and computational goals. The proposed MBPT-based framework targets semiconducting photocathodes, where electron-electron scattering mechanisms are negligible. While electron-phonon scattering is the primary collision mechanism in semiconductors, in alkali antimonides it leads to electronic band renormalizations of the order of 100~meV~\cite{gupt+17jap}, which corresponds to less than 1\% of the expected emission yield~\cite{anto+20prb}. Since the primary goal of the current implementation is to predict the spectral dependence of the photoemission yield, which is predominantly governed by the energy-dependent probability of excitation (Step 1) and emission (Step 3 discussed below), treating the transport step as purely elastic significantly reduces computational costs while retaining high predictive accuracy for ultra-thin films.

To quantify the error introduced by neglecting inelastic scattering events during transport, we estimate the electron survival probability using the attenuation length $\lambda \approx26$~nm identified experimentally for low-energy electrons in Cs$_3$Sb~\cite{gald+20jcp}. Assuming a uniform photoexcitation profile across a thin film of thickness $d$, the fraction of electrons reaching the surface without an inelastic collision scales according to the Beer-Lambert law. The average escape probability is obtained as:
\begin{equation}
    \langle P \rangle = \dfrac{1}{d} \int_0^d e^{-x/\lambda}dx = \dfrac{\lambda}{d} (1-e^{-d/\lambda}).
\end{equation}
For the sub-10-nm film thicknesses of our experimental benchmarks (e.g., $d = 9$~nm for the $\text{Cs}_3\text{Sb}$ sample probed by Parzyck \textit{et al.}~\cite{parz+22prl}), this expression yields a pristine survival rate for ballistic transport of almost 85\%. This result shows that the overestimation of the QE introduced due to neglecting inelastic scattering pathways is capped at an error margin around or below 15\%, which justifies the elastic approximation for these ultrathin configurations.

In this framework, the QE is determined by the product of the excitation and emission probability, 
\begin{equation}
    \text{QE}(\omega) \propto P_{\text{Excitation}}(\omega) \cdot P_{\text{Emission}}(\omega),
    \label{eq:QE}
\end{equation}
consistent with the mathematical simplification adopted in the Dowell-Schmerge photoemission model~\cite{dowe-schm09prab}. Future extensions of the current development may lift this simplification and include scattering probabilities computed, for example, from Monte-Carlo simulations~\cite{kark+13jap,gupt+17jap,chub+21jap}.

\subsection{Step 3: Emission Probability}
The emission probability quantifies the fraction of charge carriers escaping the surface. Necessary ingredients for this assessment are the accurate determination of the vacuum potential $V_0$, either from the experimental work function or from DFT calculations on surface slabs~\cite{anto+20prb,schi+22prm}, and the potential barrier. Here, we adjust the onset to the measured work function and assume an idealized barrier described by a step function:
\begin{equation}
T(E) = \begin{cases}
\frac{4 \sqrt{E (E - V_0)}}{(\sqrt{E} + \sqrt{E - V_0})^2} & \text{for } E \geq V_0 \\ \\
0 & \text{for } E < V_0.
\end{cases}
\end{equation}
The transmission function $T(E)$ adopted for hexagonal K$_3$Sb is shown in Fig.~\ref{fig:plots_K3Sb}(b), where $V_0 = 2.9$~eV matches the experimental value~\cite{spic58pr}. Below this threshold, no transmission is possible, while for $E \geq V_0$, $T(E)$ grows steeply and asymptotically according to
\begin{equation}
    \lim_{E\rightarrow \infty} T(E) = 1.
\end{equation}

The total emission probability, $P_{\text{Emission}}(\omega)$, is defined by convoluting $T(E)$ with the energy distribution function, $D_{\lambda}(E)$, of the excited electron generated from the $\lambda$-th excitation. $D_{\lambda}(E)$ represents the normalized probability of finding the photogenerated electron at the energy $E$ and is determined by summing the \textbf{k}-resolved electron weights computed from the BSE (Eq.~\ref{eq:w_e}) targeting QP states above the band gap:
\begin{equation}
    D_{\lambda}(E) = \sum_{c \, = \, n_{\text{CBm}}}^{n_{\text{CBmax}}} \sum_{\mathbf{k}}w_{c\mathbf{k}}^{\lambda} \, \delta(E - \varepsilon_{c\mathbf{k}}^{QP}).
    \label{eq:D}
    \end{equation}
In Eq.~\eqref{eq:D}, the first sum runs from the lowest conduction band ($n_{\text{CBm}}$) to the uppermost unoccupied state included in the solution of the BSE ($n_{\text{CBmax}}$). In the implemented workflow, QP energies are assigned via a rigid scissor shift corresponding to the QP correction to the fundamental gap from DFT. This numerical shortcut is validated by full $G_0W_0$ calculations (Fig.~S2) and physically justified by $T(E)$ being most sensitive to energy corrections near the emission threshold $V_0$. 

An example of $D_{\lambda}(E)$ computed for the excitation at 3.6~eV in hexagonal K$_3$Sb is shown in Fig.~\ref{fig:plots_K3Sb}(c). The $\delta$-like shape of this function (Eq.~\ref{eq:D}) is reflected in the plot, displaying a charge-carrier distribution between 2.9 and 3.4~eV. Additional contributions at lower energies are non-zero but so much smaller in magnitude that they are not visible in Fig.~\ref{fig:plots_K3Sb}(c).

For each excitation $\lambda$ computed from diagonalizing Eq.~\eqref{eq:BSE} the associated emission probability defined as:
\begin{equation}
P_{\text{Emission}}^{\lambda} = \int_{E_{\text{CBm}}}^{E_{\text{max}}} T(E)D_{\lambda}(E)dE .
\label{eq:P_emission}
\end{equation}
In Eq.~\eqref{eq:P_emission}, the lower integration boundary is the conduction band minimum ($E_{\text{CBm}}$) while the upper limit ($E_{\text{max}}$) corresponds to the available energy range from the solution of the BSE. These limits are the energy values related to $n_{\text{CBm}}$ and $n_{\text{CBmax}}$ defined in Eq.~\eqref{eq:D}. To compute $P_{\text{Emission}}(\omega)$ entering Eq.~\eqref{eq:QE}, $P_{\text{Emission}}^{\lambda}$ must be weighted by the contribution of the $\lambda$-th exciton to the total spectrum $\mathcal{S}(\omega) \equiv \Im \epsilon_M (\omega)$ [see Eq.~\eqref{eq:epsilon_Lorentz}].
To this end, we introduce the exciton-selective excitation probability
\begin{equation}
p_{\lambda} (\omega) = \dfrac{\mathcal{S}_{\lambda} (\omega)}{\mathcal{S} (\omega)},
\end{equation}
where 
\begin{equation}
\mathcal{S}_{\lambda}(\omega) = \frac{4 \pi^2}{\pi \Gamma \Omega} |\mathbf{t}_{\lambda}|^2 \left(\frac{\Gamma^2}{(\omega - E_{\lambda})^2 + \Gamma^2} - \frac{\Gamma^2}{(\omega + E_{\lambda})^2 + \Gamma^2} \right),
\end{equation}
to calculate the overall emission probability
\begin{equation}
P_{\text{Emission}} (\omega) = \sum_{\lambda} p_{\lambda} (\omega) P_{\text{Emission}}^{\lambda},
\end{equation}
representing the probability that an absorbed photon with frequency $\omega$ generates an emitted electron.

$P_{\text{Emission}} (\omega)$ computed for hexagonal K$_3$Sb is visualized in Fig.~\ref{fig:plots_K3Sb}(d). The steep onset around 2.9~eV [Fig.~\ref{fig:plots_K3Sb}(b)] stems from an ideally flat, step-like barrier at the vacuum interface and is modulated immediately above threshold ($E>$~3.0~eV). The monotonic increase persists up to about 4.0~eV, when the slope of the emission yield decreases drastically compared to the lower-energy region ($3.0 \leq E \leq 4.0$).  The explicit inclusion of surface roughness in the photoemission model, as proposed by several independent studies in the literature~\cite{kark-baza15prap,dimi+17jap,gevo+18prab,jens+19jap,huan+19prab,saha+23jap}, would lower the threshold by 0.1-0.2~eV, smoothing the onset. Under these assumptions, the QE is computed from the convolution of $P_{\text{Emission}}(\omega)$ with the excitation probability derived from the imaginary part of the macroscopic dielectric function (Eq.~\ref{eq:epsilon_Lorentz}). This leads to the modulated probability that a photon with frequency $\omega$ creates an exciton, which, after dissociation, generates an emitted electron (Eq.~\ref{eq:QE}):
\begin{equation}
P_{\text{Yield}} (\omega) = P_{\text{Excitation}} (\omega) \cdot P_{\text{Emission}}(\omega) .
\label{eq:P_yield}
\end{equation}

The computing power requested to run these simulations is dictated by the costs of the MBPT runs. As showcased in Table~S1 for CsK$_2$Sb and Na$_2$KSb, the $G_0W_0$/BSE simulations require most time and resources, while the final post-processing for the three-step photoemission model is highly efficient, executing in a few minutes on a standard local workstation.

\section{Model Validation: Alkali Antimonide Photocathodes}\label{sec:hex}
We validate the developed three-step model from \textit{ab initio} many-body theory against experimental photoemission data available in the literature. In this comparison, we align the simulated QE to the maximum experimental value. This empirical step is necessary to connect the microscopic description provided by our \textit{ab initio} model to measurements performed on films with a finite thickness. Here, we focus on the qualitative description of photoemission yield, adopting quantitative refinements for comparison with our experimental data on Cs$_3$Sb (see Sec.~\ref{sec:Cs3Sb}).

\begin{figure}
	\centering
	\includegraphics[width=0.5\textwidth]{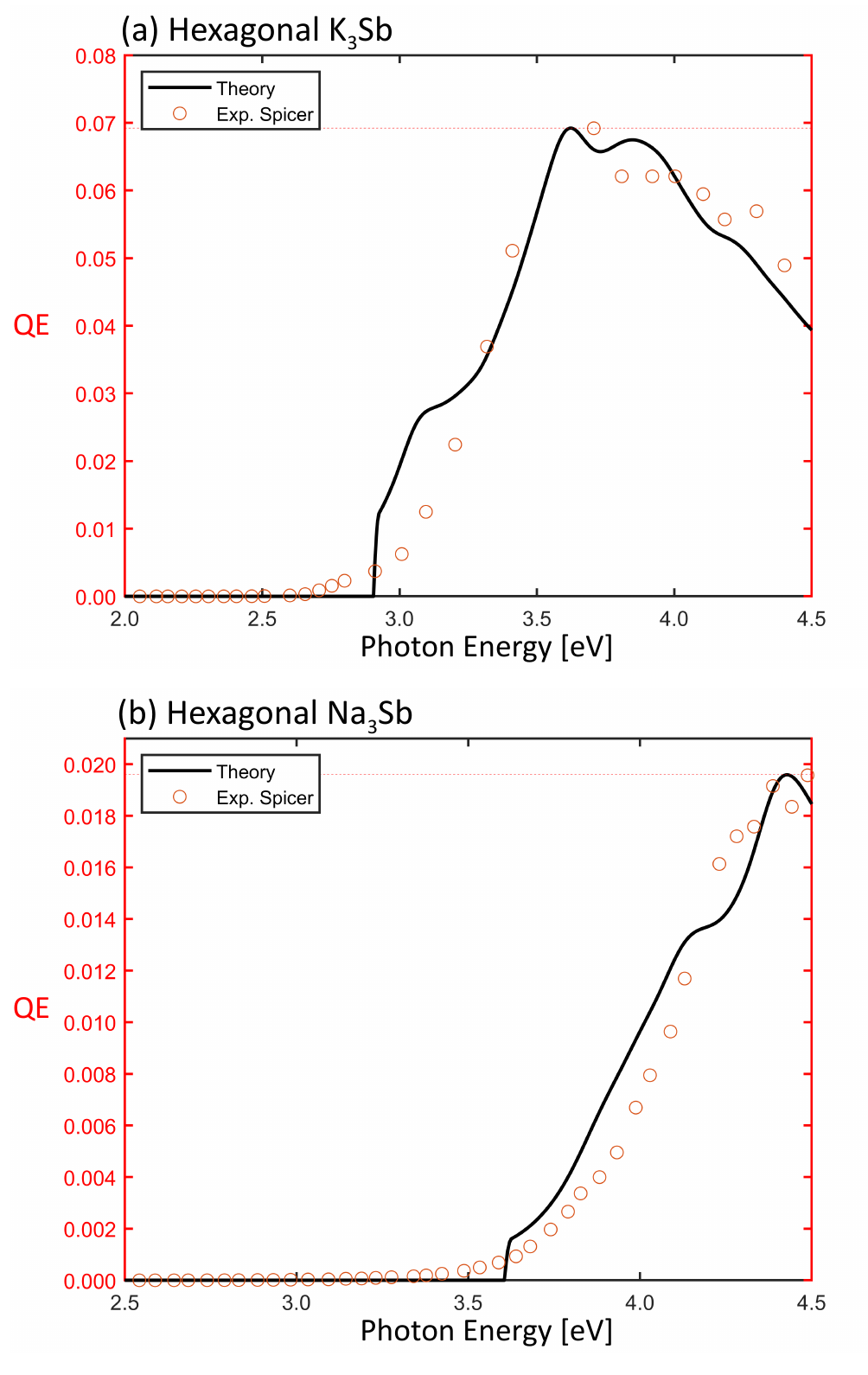}
\caption{Quantum emission computed for the hexagonal phase of (a) K$_3$Sb and (b) Na$_3$Sb (solid lines) with a Lorentzian broadening of 100~meV, compared against experimental data from Spicer~\cite{spic58pr} (empty dots). The dotted horizontal lines indicate the QE value used for aligning the \textit{ab initio} many-body prediction to the experimental data.
}
	\label{fig:results_binary}
\end{figure}

We start with hexagonal K$_3$Sb, a semiconducting photocathode measured by Spicer in 1958~\cite{spic58pr} and used in Sec.~\ref{sec:model} to illustrate the implemented steps in the model. This material exhibits a rich optical absorption spectrum in the visible region~\cite{schi-cocc25prm}, giving rise to a modulated photoemission curve that is fully captured by our \textit{ab initio} simulations [Fig.~\ref{fig:results_binary}(a)]: The steep increase of the QE up to its maximum value of 0.07 reached at 3.6~eV is followed by a smoother decrease above 4.0~eV, due to the reduction of the optical absorption in this energy region~\cite{schi-cocc25prm}. Furthermore, between 3.0 and 3.2~eV, we notice a localized difference in the spectral trends: the simulation exhibits a distinct shoulder-like plateau that is absent in the experiment. This feature is intrinsic to the calculated many-body joint density of states at 0~K (see Fig.~\ref{fig:plots_K3Sb}). In Spicer's measurement~\cite{spic58pr}, this step-like variation is smoothed out by thermal broadening and instrumental bandwidth.

The most noticeable difference between the experimental data and our computational prediction is at the onset, where the model predicts a much sharper transition than the measurement. This characteristic is a direct consequence of describing the emission barrier as a step function [Fig.~\ref{fig:plots_K3Sb}(b)]. Moreover, as already noted by Spicer~\cite{spic58pr,spic60jap}, photoemission from K$_3$Sb is significantly influenced by impurities, leading to a non-zero smearing of the QE near the onset. Since our \textit{ab initio} calculations are performed for an ideal K$_3$Sb bulk crystal~\cite{schi-cocc25prm}, these physical effects are not included. It is also worth noting that the empirical version of Spicer's model, which could successfully describe the photoemission yield of cubic alkali antimonide crystals~\cite{spic58pr}, is unable to reproduce the experimental QE of K$_3$Sb~\cite{spic58pr,spic60jap}, due to the complex band structure stemming from its hexagonal lattice. The overall very good performance of our model demonstrates the need for an accurate description of the electronic structure to properly simulate the photoemission yield of this material. Future refinements, including defects and impurities, are expected to provide an even closer agreement with the measurements.

Next, we compare the simulated and measured QE for another hexagonal alkali antimonide, namely Na$_3$Sb [Fig.~\ref{fig:results_binary}(b)]. This material exhibits a high threshold for intrinsic photoemission above 3.5~eV, given by its band gap of approximately 1.1~eV and its electron affinity around 2.4~eV~\cite{spic58pr,spic60jap}. After a steep increase, the QE reaches its maximum of 0.02 at about 4.4~eV. This behavior is well reproduced by our \textit{ab initio} many-body model, with the computed QE curve modulated by the underlying optical absorption~\cite{schi-cocc25prm}, see Fig.~S3(b). Similar to the case of K$_3$Sb shown in Fig.~\ref{fig:results_binary}a, the simulated curve for Na$_3$Sb displays a step-like plateau between 4.1 and 4.3~eV, which appears as a continuous, uniform slope in the experimental data (Fig.~\ref{fig:results_binary}b). This difference is attributed to the natural chemical and structural disorder in the samples that is not accounted for in our calculations of idealized single crystals. At low energies, the experimental photoemission yield of Na$_3$Sb is affected by impurities~\cite{spic58pr,spic60jap}, leading to non-zero values below the onset and to a smoother increase compared to our simulation of the ideal crystal and the emission barrier approximated by a step-function. Similar to K$_3$Sb, Spicer's empirical model is incapable of fitting the measurement, due to the non-trivial band structure of hexagonal Na$_3$Sb.

\section{Photoemission Predictions for Ternary Alkali Antimonides}
After the successful benchmark of our \textit{ab initio} many-body method for calculating photoemission yield against experimental data available from the literature for binary alkali antimonides, we put it to the test for ternary compositions (Na$_2$KSb and CsK$_2$Sb), representing state-of-the-art semiconducting materials employed for photocathodes in particle accelerators~\cite{musu+18nimpra,scha+23jmcc,moha+23micromachines}. For this comparison, we take as a reference our spectra measured at HZB.

\begin{figure}
	\centering
	\includegraphics[width=0.5\textwidth]{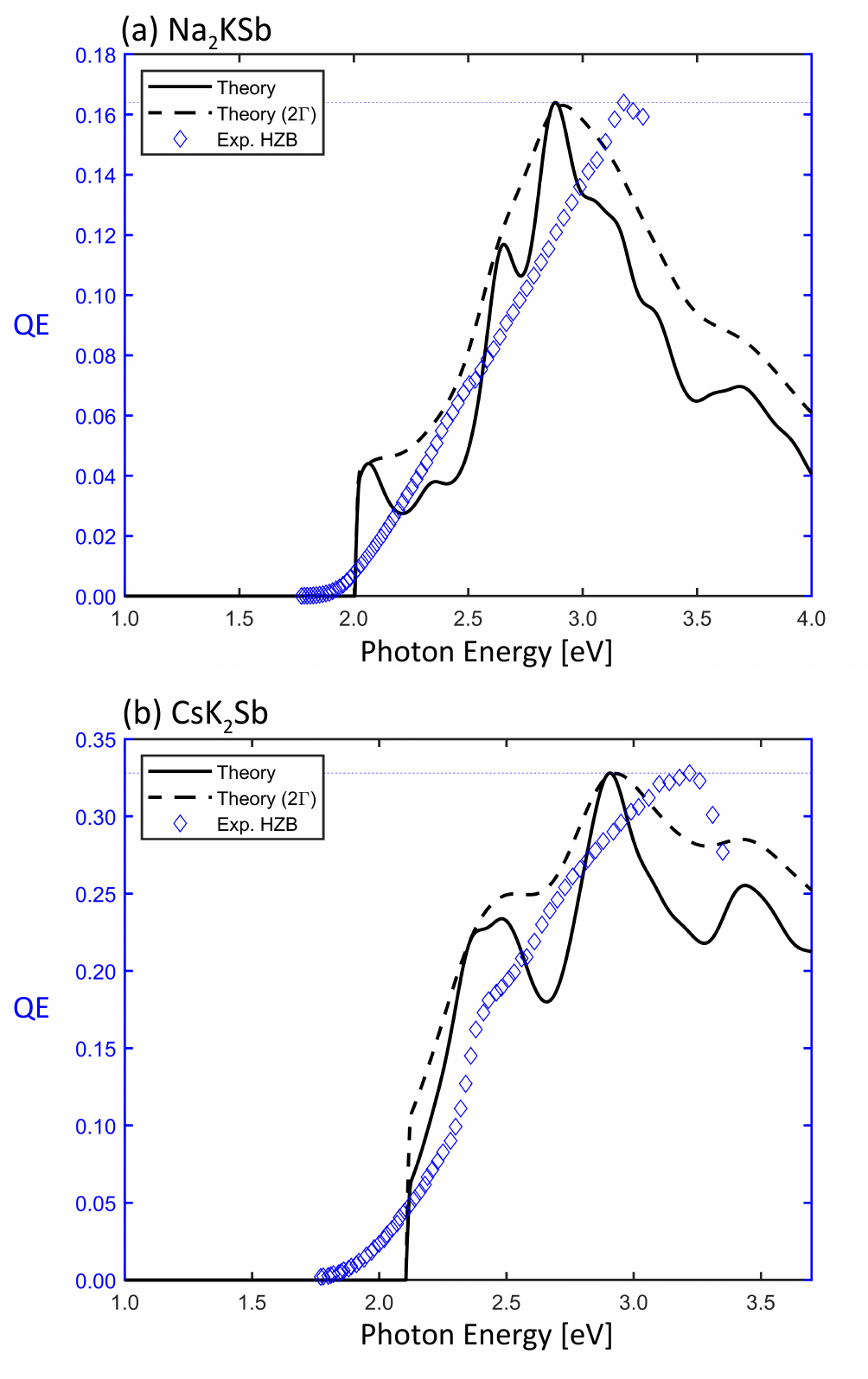}
\caption{QE of cubic (a) Na$_2$KSb and (b) CsK$_2$Sb crystals measured at HZB (empty diamonds) and computed from the proposed \textit{ab initio} many-body photoemission model using a broadening $\Gamma = 100$~meV (solid lines) and $2\Gamma = 200$~meV (dashed lines). The dotted horizontal bars indicate the QE value used to align the calculated spectrum to the experimental data.
}
	\label{fig:results_ternary}
\end{figure}

Na$_2$KSb is known for its stability~\cite{erja96ass,sant+24jpm,yue+22prb} and QE~$>10\%$ in the visible-to-infrared range~\cite{cult+16apl}, as confirmed by our measurements [Fig.~\ref{fig:results_ternary}(a)]. The investigated sample exhibits an emission threshold around 2.0~eV, with the QE rising steeply toward a maximum at 3.2~eV. 
The quantum yield computed from first principles reproduces well the experimental trend. As discussed for the binary compounds above, the wealth of details in the simulated QE stems directly from the optical absorption spectrum calculated from GW+BSE~\cite{amad+21jpcm}, see Fig.~S4(a). Given the relatively low vacuum potential of this material [Fig.~S1(d)], most optical excitations in Na$_2$KSb effectively modulate the emission probability [Fig.~S4(b)] and, ultimately, the final QE prediction [Fig.~\ref{fig:results_ternary}(a)].  

To assess the impact of microstructural and morphological factors (e.g., polycrystallinity, grain boundaries, and local chemical disorder) that are absent in our calculations on pristine crystals, we show an alternative QE prediction curve computed with an increased broadening parameter ($2\Gamma=200$~meV, dashed black line). In the case of Na$_2$KSb [Fig.~\ref{fig:results_ternary}(a)], besides the steep onset inherited from the step-like potential barrier included in our model and a rigid red-shift of the main maximum by about 100~meV, the broadened curve follows the slope of the experimental curve, matching its few discernible features, above all, the ``kink'' around 2.5~eV.

The QE of CsK$_2$Sb, one of the most popular bialkali antimonides used for photocathodes in particle accelerators~\cite{schm+18prab,musu+18nimpra,scha+23jmcc}, assumes values between 3\% and 35\% across the entire visible range above threshold ($\sim$2.0~eV), see Fig.~\ref{fig:results_ternary}(b). Our theoretical prediction follows the same spectral modulation exhibited by the  measurement~\cite{schm+18prab}, which, however, is characterized by a large broadening due to structural and morphological disorder in the sample. While the QE predicted with $\Gamma=100$~meV enables a clear identification of the main spectral features and their connection to optical absorption maxima [Fig.~S4(c)], doubling the smearing parameter ($\Gamma=200$~meV) leads to a better match with the experimental curve. 

After a smooth onset between 2.0 and 2.2~eV, which is missed in the calculation due to the chosen step-like potential barrier, the measured QE reaches a local maximum at $\sim$2.4~eV. This spectral feature, including the subsequent change of slope, is qualitatively reproduced by the \textit{ab initio} model [Fig.~\ref{fig:results_ternary}(b)]. At higher energies, the simulation with the larger broadening (dashed line) washes out the local minimum present at 2.6~eV in the pristine spectrum, generating a smooth profile that better tracks the continuous upward experimental trend up to the absolute yield maximum (at 3.2~eV in experiment and 2.9~eV in the \textit{ab initio} simulation) before reproducing the subsequent sudden drop in QE [Fig.~\ref{fig:results_ternary}(b)].

The systematic energetic red-shift of about 200~meV between the calculated peaks and the experimental thresholds can be ascribed to the convergence limits expected for the GW/BSE calculations targeting high conduction states ($E> 1.5$~eV above the CBM) that have been truncated in the solution of the BSE~\cite{cocc+19jpcm}. Numerical refinements of this underlying description of optical absorption of CsK$_2$Sb, including a broader range of unoccupied states, are expected to substantially improve the agreement with the measured QE in the high-energy region.

\section{Quantitative QE Predictions for Cesium Antimonide} \label{sec:Cs3Sb}
In the results discussed so far, the \textit{ab initio} predicted QE was manually aligned to the maximum of the measured curve (see Fig.~\ref{fig:results_binary} and Fig.~\ref{fig:results_ternary}). While this approach is fully reasonable for validating the method and obtaining a qualitative comparison of the main spectral features, it has limited predictive power under realistic conditions, where independent measurements may deliver different results, or when there is no reference to compare with. In the following, we examine the photoemission yield of cubic Cs$_3$Sb, likely the most popular alkali antimonide photocathode material~\cite{musu+18nimpra,scha+23jmcc,gald+21apl}. We compare our \textit{ab initio} results with four experimental datasets, representing a diverse range of growth techniques and sample qualities: Spicer's pioneering measurements~\cite{spic58pr}, recent benchmarks from Karkare's group at Arizona State University (ASU)~\cite{saha+22apl}, data collected at Cornell University~\cite{parz+22prl}, and measurements performed at HZB~\cite{schm19thesis}.

\begin{figure}
	\centering
	\includegraphics[width=0.5\textwidth]{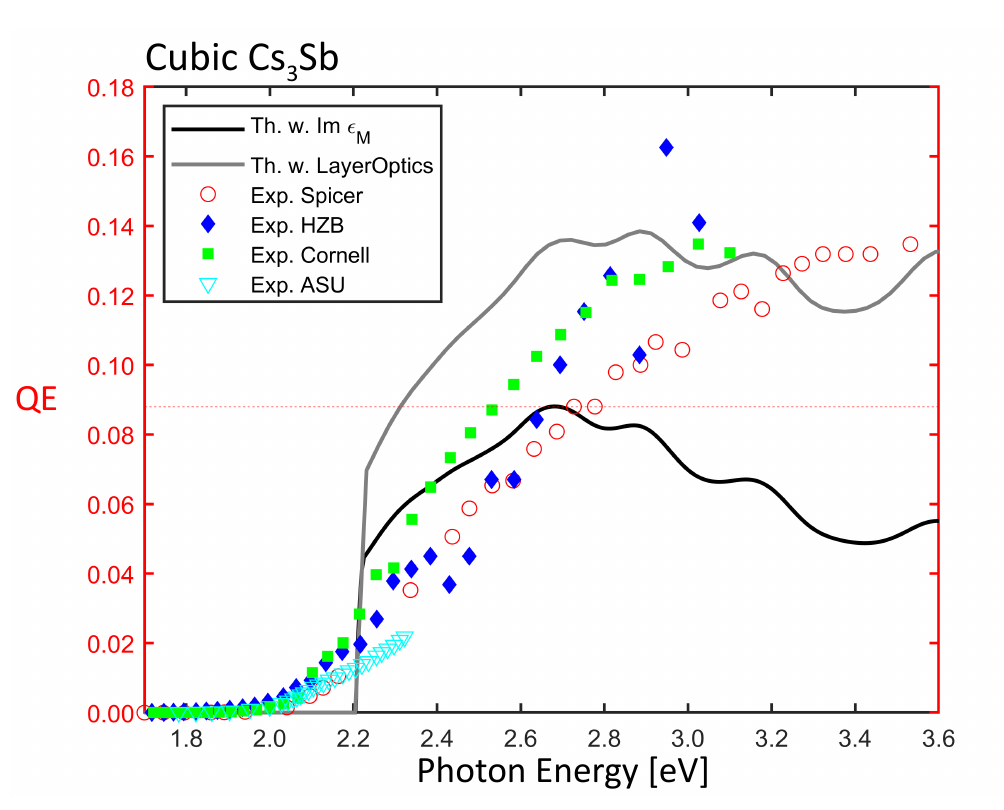}
\caption{Quantum emission curve computed for cubic Cs$_3$Sb using a broadening of 100~meV and compared against experimental data from Spicer~\cite{spic58pr}, HZB~\cite{schm19thesis}, Cornell University~\cite{parz+22prl} and Arizona State University (ASU)~\cite{saha+22apl}. The dotted bars indicate the QE value used to align the \textit{ab initio} many-body prediction with the experimental data. A quantitative prediction for a 9~nm-thick thin film calculated with \texttt{LayerOptics} is provided without manual alignment of the QE.
}
	\label{fig:results_Cs3Sb}
\end{figure}

All the experimental results presented in Fig.~\ref{fig:results_Cs3Sb} exhibit a smooth onset around 2.0~eV. While the measurements from ASU extend only up to 2.5~eV~\cite{saha+22apl}, the remaining three datasets follow a similar energy-dependent trend up to approximately 2.7~eV. Between 2.3~eV and 2.7~eV, the Cornell sample shows a steep increase, reaching a maximum QE of about 13\% close to 3.0~eV. The highest emission yield is scored by the cathode grown at HZB (QE~$\sim$17\% at 2.95~eV), while Spicer's measurements follow a slower monotonic increase, culminating in a maximum QE of approximately 14\% at 3.5~eV.

A common feature of the simulated results is a sharper onset compared to the experimental curves. This can be attributed to several factors inherent to real-world photocathodes. First, the inclusion of an ideally flat potential barrier eliminates complex surface mechanisms that rule the response in the near-threshold region (i.e., $\sim$0.1-0.2~eV above the onset)~\cite{huan+19prab,saha+23jap}, creating the idealized step-like potential barrier (Fig.~\ref{fig:plots_K3Sb}b) that is reflected in the sharp onset of QE curves. Alkali antimonide surfaces are characterized by chemical roughness, inducing spatial variations in the work function on the order of 0.3–0.4~eV~\cite{gald+21apl}. When combined with the typical monochromator bandwidth of 5–10~nm used in the measurements, these effects naturally broaden and soften the sharp excitonic features and step-like thresholds predicted by our theory at 0~K. On the other hand, the variations among the datasets shown in Fig.~\ref{fig:results_Cs3Sb} can be attributed to the growth method. While traditional co-deposited films frequently exhibit a ``knee'' feature around 2.1~eV followed by a defect-driven tail~\cite{woot+73jap}, quasi-epitaxial films, such as those grown on SrTiO$_3$ at ASU~\cite{saha+22apl} or via deposition-recrystallization at HZB~\cite{schm19thesis}, show a much more gradual increase in QE. These quasi-epitaxial samples are typically ultrathin (10–15 nm) and more closely resemble the ideal crystal structure used in the many-body calculations. 

Finally, it is worth noting that, while the traditional photoemission threshold for Cs$_3$Sb is often reported at about 2.1~eV, recent photoemission electron spectroscopy measurements have suggested a significantly lower work function of about 1.5~eV~\cite{kach+23apl}. In this context, the threshold features in spectral response curves may not solely represent the vacuum barrier, but rather specific structures in the density of states or excitonic transitions. This view is supported by the detailed energy distribution functions derived from our BSE calculations~\cite{amad+21jpcm,cocc+19jpcm,schi-cocc25prm}.

The large discrepancies among the measured QE curves shown in Fig.~\ref{fig:results_Cs3Sb} further motivate the need for unbiased \textit{ab initio} predictions exploring idealized pristine materials at 0~K. At the same time, this variety of experimental data, reflecting different sample preparation and measurement conditions, challenges an empirical alignment of the computed photoemission yield with the experimental benchmarks. 

 In Fig.~\ref{fig:results_Cs3Sb}, the manual alignment between the \textit{ab initio} result (black curve) and the measurements (colored dots) was performed in the region between 2.5 and 2.7~eV, where the results from Spicer, HZB, and Cornell are in closest proximity. However, since this window is just above threshold, it exacerbates the visual mismatch across the rest of the spectrum. Below 2.4~eV, the step-like potential barrier chosen in our model inhibits a faithful description of the experimental onset compared to all datasets. Conversely, at higher energies, the calculated curve fails to capture the steep QE increase featured by the Cornell sample, remaining closer to Spicer's and HZB data up to 2.7~eV without, however, displaying a fully satisfactory line match.
The region between 2.7 and 3.6~eV is the most problematic for the theoretical prediction, as all experimental curves are underestimated by at least a factor of 2 (Fig.~\ref{fig:results_Cs3Sb}). This discrepancy can be ascribed to the large number of electronic bands involved in such high-energy transitions, which were not fully included in the BSE calculation originally performed to investigate the optical absorption of Cs$_3$Sb~\cite{cocc+19jpcm}. 

The deviations discussed above motivate us to enhance the predictive power of our model by post-processing the dielectric function with Fresnel equations, incorporating macroscopic parameters such as the film thickness, angle of incidence, and light polarization. This procedure is implemented in the \texttt{Python} package \texttt{LayerOptics}~\cite{vorw+16cpc}, a tool designed to handle complex optical geometries by rotating the dielectric tensor to match specific surface orientations. While the cubic symmetry of Cs$_3$Sb does not require the advanced features of \texttt{LayerOptics} developed to treat highly anisotropic materials~\cite{cocc+16prb,cocc+18pccp}, this post-processing allows us to assume normal incidence for the incoming radiation and to account for a film thickness $d=9$~nm, matching the specifications of the sample grown at Cornell~\cite{parz+22prl}.

The excitation probability assumed in the Maxwell formalism of \texttt{LayerOptics} is defined as:
\begin{equation}
\tilde{P}_{\text{Excitation}}(\omega) = 1 - (R (\omega) + T (\omega)) ,
\label{eq:P_Exc_quant}
\end{equation}
where $R (\omega)$ and $T (\omega)$ are the calculated reflectance and transmittance, respectively. By substituting $\tilde{P}_{\text{Excitation}}$ into the expression for the photoemission yield, we obtain:
\begin{equation}
\tilde{P}_{\text{Yield}}(\omega) = \tilde{P}_{\text{Excitation}}(\omega) \cdot P_{\text{Emission}}(\omega) .
\label{eq:P_yield_quant}
\end{equation}
This result is the quantitative counterpart of Eq.~\eqref{eq:P_yield}, and enables a direct comparison with experimental data without manual adjustments.

The QE of Cs$_3$Sb predicted with \texttt{LayerOptics} is in much better agreement with the experimental data (gray curve in Fig.~\ref{fig:results_Cs3Sb}). The maximum featured at approximately 14\% is remarkably close to both the Cornell and HZB measurements. Notably, post-processing with \texttt{LayerOptics} notably matches the prediction of Spicer's QE between 3.0 and 3.5~eV. Beyond the quantitative improvement, this post-processing also enhances the qualitative spectral behavior. While the low-energy portion of the curve is essentially translated to higher QE values, the range between 2.7 and 3.2~eV no longer exhibits the systematic envelope reduction that caused the mismatch between the measurements and the unprocessed \textit{ab initio} results. In particular, the theoretical curve obtained with \texttt{LayerOptics} closely follows the profile recorded at Cornell and better replicates the spectral increase observed in HZB and Spicer's datasets, despite the persistent overestimation in the 2.2–2.7~eV range. 

This improved description demonstrates that accounting for macroscopic optical effects, specifically the interference and polarization-dependent reflectance within the thin film, is critical for bridging microscopic many-body theory with realistic photoemission observables. By incorporating the experimental geometry and film thickness, we move from a qualitative spectral analysis to a truly predictive quantitative framework.

\section{Summary and Conclusions}
In summary, we presented an \textit{ab initio} many-body extension of the three-step photoemission model for semiconducting photocathodes, directly integrating the $GW$ approximation and the solution of the BSE on top of DFT. This methodology successfully links the microscopic electronic structure of the materials, including quasiparticle and excitonic effects, with the QE. We validated our model against experimental data for binary and ternary alkali antimonides, demonstrating very good agreement with measurements obtained from independent groups. Our predictions offer in-depth insight into the spectral response of the crystals, capturing complex features, such as electronic self-energy and excitonic transitions, that are intrinsic to photo-excitation and cannot be captured by DFT alone.
The transition from a qualitative spectral analysis to a quantitative prediction of the QE allowed us to move beyond empirical alignment, yielding quantitative predictions that accurately match absolute QE values with remarkable accuracy. This demonstrates that the combination of many-body perturbation theory and classical Fresnel optics provides a complete framework for describing the photoemission process.

In conclusion, this work successfully bridges the gap between an accurate microscopic characterization of the spectral response of semiconductors and the macroscopic photoemission process. The integration of MBPT methods overcomes the accuracy limitations of previous models based on semi-local DFT and provides a robust, parameter-free tool for predicting the spectral characteristics of semiconducting photocathodes, opening the path for the rational design and optimization of next-generation electron sources.

Future work will focus on quantitative improvements, such as incorporating scattering contributions via Monte Carlo methods and providing more realistic descriptions of potential barriers. More broadly, this framework establishes a clear roadmap for how theory and experiment can actively converge to bridge the remaining gaps. From the modeling side, we aim to bring our simulations closer to real samples by explicitly incorporating surface adsorbates, oxidation layers, and native point defects within the crystal lattice. From the experimental side, our parameter-free model stimulates refinements in the growth and characterization procedures. Transitioning away from conventional (co-)deposition techniques and toward single-crystal epitaxial growth will minimize grain-boundary scattering and microstructural disorder, bringing laboratory line shapes closer to our pristine crystalline profiles. Furthermore, by explicitly tracking the $\mathbf{k}$-dependent electronic states involved in optical absorption and subsequent emission, our approach inherently contains all necessary information to evaluate other critical observables, such as the mean transverse energy and thermal emittance, going beyond semi-empirical models proposed in the literature~\cite{saha+23jap}. By providing an exemplary bridge between atomistic quantum theory and state-of-the-art experiments,  this work represents a milestone in photocathode development.

\section*{Acknowledgments}
The authors thank Sonal Mistry and Martin Schmei{\ss}er for valuable discussions. This work was funded by the German Research Foundation (DFG), Project No. 490940284. Computational resources were provided by the High-Performance Computing Centers at the University of Oldenburg (clusters CARL and ROSA), funded by the Lower Saxony Ministry for Science and Culture and by the DFG (Project Nr. INST 184/157-1 FUGG and INST 184/225-1 FUGG, respectively) and at the Friedrich-Schiller University Jena (cluster ARA).

\section*{Author Contributions}
R.S. and C.C. developed the theoretical framework. R.S. implemented the computational method, performed all \textit{ab initio} simulations, prepared the figures, curated the data, and wrote the original manuscript draft. C.C. conceptualized the work, validated the results, and finalized the manuscript. J.D., C.W., J.K., and T.K. designed the experiments and developed the apparatus for spectral response. J.D., C.W., and J.K. performed the measurements. J.D., C.W., T.K., and A.G. analyzed and interpreted the experimental data. C.C. and T.K. acquired the resources. All authors contributed to the discussion of the results, provided critical feedback, and reviewed the final manuscript.

\section*{Data availability}
The data that support the findings of this article are openly available in Zenodo. DOI: 10.5281/zenodo.18656520
%

%

\end{document}